\documentclass[fleqn,usenatbib]{mnras}

\usepackage{newtxtext,newtxmath}

\usepackage[T1]{fontenc}

\DeclareRobustCommand{\VAN}[3]{#2}
\let\VANthebibliography\thebibliography
\def\thebibliography{\DeclareRobustCommand{\VAN}[3]{##3}\VANthebibliography}

\usepackage{graphicx}	
\usepackage{amsmath}	
\usepackage{multirow} 
\usepackage{comment}

\title[AGN Properties Through a Changing-look Event]{Tracing Active Galactic Nuclei Properties Through a Changing-look Event}

\author[J. Carpenter, S. Raimundo, C. R. Angus, K. Auchettl]{
Joel Carpenter,$^{1}$\thanks{E-mail: j.carpenter@soton.ac.uk (UOS)}
Sandra Raimundo,$^{1,2}$ 
Charlotte Angus,$^{3}$ 
Katie Auchettl$^{4,5}$ 
\\
$^{1}$School of Physics and Astronomy, University of Southampton, Southampton SO17 1BJ, UK\\
$^{2}$DARK, Niels Bohr Institute, University of Copenhagen, Jagtvej 155, Copenhagen N, 2200, Denmark\\
$^{3}$Astrophysics Research Centre, School of Mathematics and Physics, Queen’s University Belfast, BT7 1NN, UK\\
$^{4}$School of Physics, The University of Melbourne, VIC 3010, Australia\\
$^{5}$Department of Astronomy and Astrophysics, University of California, Santa Cruz, CA 95064, USA\\
}

\date{Accepted XXX. Received YYY; in original form ZZZ}

\pubyear{\the\year{}}

\begin{document}
\label{firstpage}
\pagerange{\pageref{firstpage}--\pageref{lastpage}}
\maketitle

\begin{abstract}
Changing-look transitions challenge our understanding of active galactic nuclei (AGN), exhibiting dramatic changes in broad-line emission and continuum flux on timescales of months to years. We present a detailed study of the spectroscopically confirmed changing-look AGN ZTF18abuamgo. Combining photometric survey data with spectroscopy spanning three epochs over 20 years, we identify a turn-on transition from a Type 1.5 to Type 1.2 AGN and estimate the timescale of this change to be as short as four years. Spectral analysis indicates that this transformation is driven by a rapid increase in accretion rate, with the Eddington ratio rising from $0.032 \pm 0.005$ in the dim state to $0.08 \pm 0.01$ in the bright state. For the first time in a changing-look AGN, we apply the Boltzmann plot method to the visible Balmer series emission, deriving broad line region electron temperatures of $11,800 \pm 900$ K and $11,900 \pm 2,400$ K in 2022 and 2024, respectively. Applying single-epoch black hole mass estimation to the brightening H$\alpha$ emission, we find a mass of $(5.0 \pm 0.4) \times 10^7 M_\odot$. The consistency in this estimate across all spectroscopic epochs suggest that even highly variable broad lines in CL-AGN do not bias the results derived using this method. Our results demonstrate that objects like ZTF18abuamgo provide a unique laboratory to study extreme AGN variability, probe the physical conditions in the broad line region, and assess the limitations of widely used black hole mass estimation methods.
\end{abstract}




\section{Introduction}

Active galactic nuclei (AGN) are among the most luminous and dynamic phenomena in the Universe, powered by accretion of matter onto supermassive black holes at the centres of galaxies. Their intense radiation, variability, and feedback into the host galaxy make them crucial laboratories for understanding both galaxy evolution and the physics of extreme environments. For decades, the unified model of AGN has provided a framework to explain the observed diversity of AGN types through orientation-dependent obscuration \citep{antonucci_unified_1993, urry_unified_1995}. Type 1 AGN display Doppler-broadened emission lines from the photoionised broad-line region (BLR), while Type 2 AGN, which lack broad lines in their spectra, were historically understood to be intrinsically similar objects viewed edge-on, with the BLR obscured by a dusty torus. Intermediate classifications (Types 1.2–1.9) have been adopted to describe sources with present but progressively weaker broad emission lines (e.g. \citealt{osterbrock_seyfert_1981-1, winkler_variability_1992, whittle_virial_1992}). 

A rare but increasingly studied subclass of AGN is the so-called changing-look AGN (CLAGN), exhibiting dramatic spectral transformations between Type 1 and Type 2 classifications over timescales of months to years. These changes are characterised by the appearance or disappearance of broad emission lines in the optical spectrum, typically accompanied by significant variations in continuum luminosity. First discovered in a handful of nearby Seyfert galaxies (e.g., \citealt{tohline_variation_1976, lamassa_discovery_2015}), CLAGN have since been identified in large-scale time-domain surveys (e.g., \citealt{macleod_systematic_2016, sheng_initial_2020, amrutha_discovering_2024}), revealing that such transitions are not as rare as once thought.

Proposed explanations for changing-look transitions broadly fall into two categories: variable obscuration and intrinsic state changes. In the former, changes in the distribution or density of dust clouds in the torus along the line of sight can obscure the central engine and partially or entirely block the BLR, resulting in suppressed or absent broad emission lines \citep{koss_bat_2017}. In the latter, the observed changes are attributed to intrinsic variations in the accretion rate onto the supermassive black hole, which lead to changes in the ionising continuum that powers the BLR. Since the BLR responds to the photoionising radiation emitted by the inner accretion disc, significant changes in the disc's structure or luminosity can result in the appearance or disappearance of broad emission lines. These "changing-state" scenarios typically invoke rapid transitions, faster than expected from viscous timescales of the accretion disc, suggesting that disc instabilities or external perturbations, such as tidal disruption events, may play a role \citep[e.g.,][]{ricci_changing-look_2023}. 

Alternative models propose analogies with spectral state transitions observed in X-ray binaries. In this framework, changes in accretion mode, such as a shift from a standard thin accretion disc to an inner advection-dominated accretion flow (ADAF), can suppress high-energy emission and reduce BLR ionisation \citep{noda_explaining_2018}. Other hypotheses include the formation or destruction of the BLR itself in response to long-term changes in accretion rate \citep{elitzur_disappearance_2009}. Recent estimates from SDSS-V suggest that CLAGN constitute approximately $0.4\%$ of the overall AGN population, with this fraction rising slightly among AGN with low Eddington ratios \citep{zeltyn_exploring_2024}. This supports the idea that low-accretion-rate systems may be more prone to structural changes in the accretion disc and BLR. 

AGN are known to exhibit variability across a wide range of timescales and wavelengths. In the optical, the continuum emission from the ultraviolet (UV)–bright accretion disc typically varies by $\sim0.2$ mag over months to years \citep{berk_ensemble_2004}. A minority of sources, roughly $10\%$, show changes exceeding 1 magnitude over decade-long baselines (e.g. extreme variability quasars; EVQs; \citep{rumbaugh_extreme_2018}, or on even shorter timescales in the case of CLAGN. At longer wavelengths, thermal mid-infrared (MIR) emission from the dusty torus generally varies by $\sim0.1$ magnitudes on year-long timescales (e.g. \citealt{kozlowski_quantifying_2010, lyu_mid-ir_2019, son_mid-infrared_2022}), with CLAGN exhibiting variations at times above $0.4$ magnitudes \citep{sheng_initial_2020}. Placing these continuum variations in context, and clarifying the physical connection between luminosity changes and the presence or absence of broad emission lines, is key to determining whether CLAGN mark the extreme tail of normal AGN variability or instead signal more fundamental changes in accretion state or obscuration. 

The physical conditions within the BLR of CLAGN remain poorly constrained. In standard AGN, the electron temperature of the BLR has been estimated using Boltzmann plots based on the hydrogen Balmer series, yielding temperatures in the range of $7,000$–$37,000$K \citep{ilic_analysis_2012}. This method uses Balmer line intensities to reflect excitation conditions, typically with a minimum of five lines (from H$\alpha$ to H$\epsilon$) to constrain the best fit \citep{popovic_balmer_2003, popovic_broad_2006}. However, this technique has not yet been applied to CLAGN, and its use is strongly constrained by the limited number of spectra with sufficiently bright and well-measured Balmer emission lines. In many cases, the low flux of higher-order transitions makes spectral fitting challenging and increases the uncertainties in the resulting Boltzmann plot, thereby reducing the reliability of the derived temperature estimates (e.g. see examples of fit quality from \citealt{mura_detailed_2007}).

Estimating the masses of distant supermassive black holes (SMBHs) is critical to many areas of galaxy and AGN evolution. Strong empirical correlations between black hole mass and galactic properties, such as bulge mass, stellar velocity dispersion, and luminosity, have been widely observed, although the physical origin of these correlations remains a matter of debate (e.g. \citealt{magorrian_demography_1998, gebhardt_relationship_2000, ferrarese_fundamental_2000, kormendy_coevolution_2013}). A widely used approach for estimating SMBH masses is the single-epoch method, which infers mass from a single optical spectrum by combining the width of broad emission lines (e.g. H$\alpha$, H$\beta$, \ion{Mg}{ii}, \ion{C}{iv}) with either the line or continuum luminosity as a proxy for the size of BLR. These estimates assume a virialised BLR, where the black hole mass can be expressed as $M_{\text{BH}} \sim R_{\text{BLR}} v^2 / G$ \citep{dibai_mass_1977, peterson_keplerian_1999} and are calibrated using the radius–luminosity (R--L) relation, an empirical scaling derived from reverberation mapping studies \citep{kaspi_reverberation_2000, peterson_central_2004, bentz_low-luminosity_2013}. This relation, combined with the measured broad-line widths, forms the basis of single-epoch black hole mass estimates \citep{vestergaard_determining_2002, vestergaard_determining_2006, greene_estimating_2005-1}, which are widely applied across large AGN samples, typically with an intrinsic scatter of $\sim0.4$–$0.5$ dex. 

However, changing-look AGN complicate this picture. Their dramatic transformations in continuum and broad-line emission raise concerns about whether single-epoch mass estimates remain reliable. While CLAGN have been shown to broadly follow the $M_{\mathrm{BH}}$–$\sigma_*$ relation \citep{jin_systematic_2022}, it remains unclear whether single-epoch estimates obtained during dim and bright states are consistent, as illustrated in the case of highly variable AGN RM160, which exhibits nonvirial BLR kinematics that challenge the virial assumption \citep{fries_sdss-v_2024}. This has large ramifications for studying SMBHs across large samples of AGN where single epoch spectra are often used to confirm their nature, for example, in the use of large surveys like the Legacy Survey of Space and Time (LSST).

In this paper, we present a detailed analysis of ZTF18abuamgo, a previously unstudied AGN at redshift $z = 0.074450$ \citep{geller_redshift_2014}, hosted in galaxy J$024545.49$–$030449.6$, which underwent a rapid optical brightening caught by the Zwicky Transient Facility (Section \ref{photometry}), with follow-up spectroscopic observations (Section \ref{spectroscopy}). The emergence of broad Hydrogen Balmer series emission lines in the recent bright phase makes ZTF18abuamgo a compelling case for studying the physical drivers of CLAGN variability. Using the multi-epoch spectroscopy, we investigate the spectral and accretion state evolution of the source (Section \ref{changing_look}). For the first time in a CLAGN, we probe the physical conditions of the broad-line region through Boltzmann plot analysis (Section \ref{blr}). We follow by testing the consistency of single-epoch black hole mass estimates before and after the changing-look transition, using both H$\alpha$ and H$\beta$ emission lines (Section \ref{bh_mass}). Finally, we discuss the interpretations of our results (Section \ref{discussion}), and summarise our findings (Section \ref{conclusion}).

\section{Observations and Data Analysis} \label{observations}

We combine photometric and spectroscopic observations of ZTF18abuamgo to study its physical properties. The photometric data show the brightness of the source as a function of time over multiple wavebands. We used photometry to identify and constrain the AGN variability. The emission lines and continuum emission present in the spectrum are used to extract information on the AGN and surrounding excitation conditions. In this section, we outline the observational data and methods used to analyse those datasets.

\begin{figure*}
	\includegraphics[width=15cm]{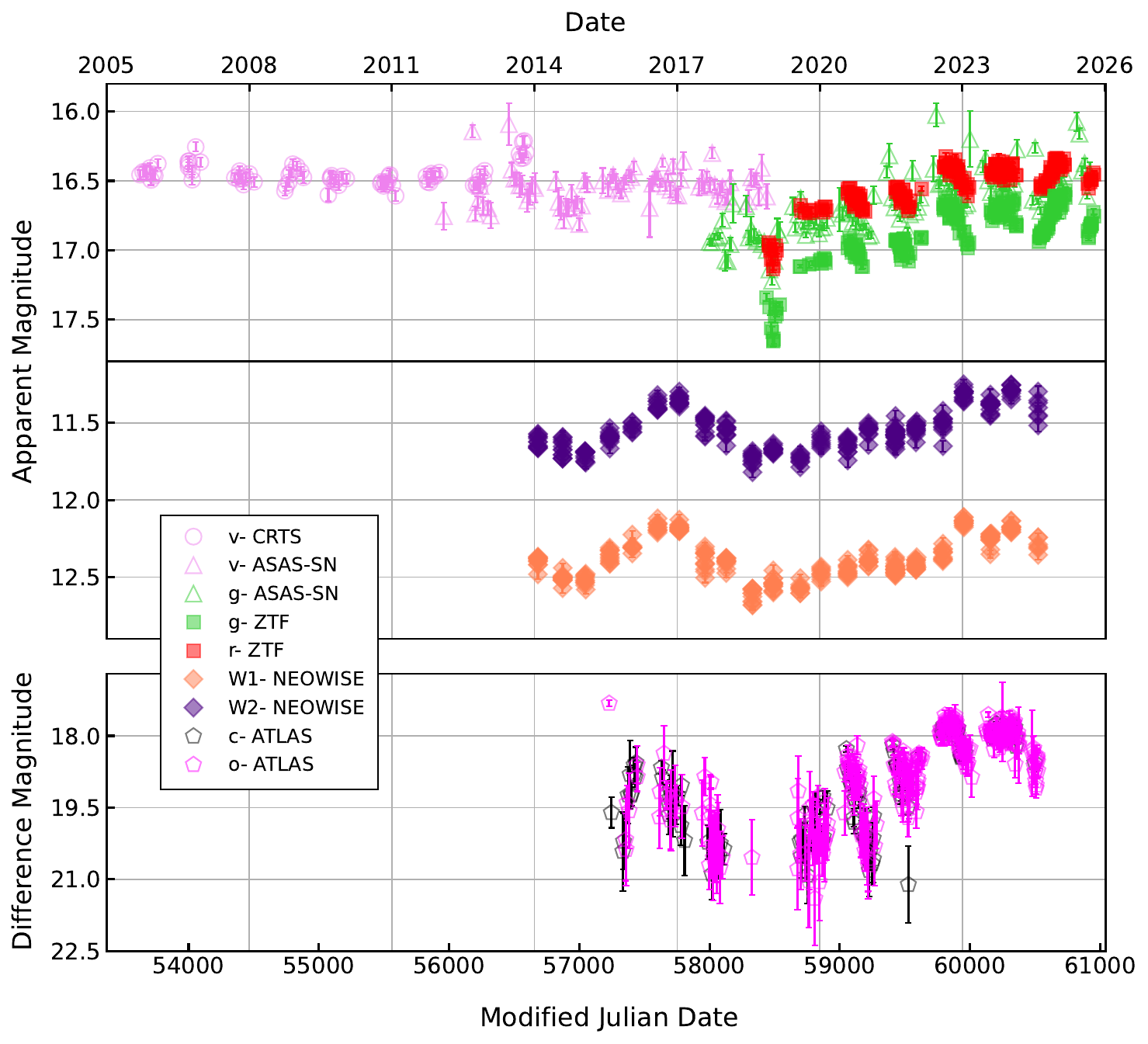}
    \caption{Photometric lightcurves in optical bands (circles, CRTS; triangles, ASAS-SN; squares, ZTF) in the top panel, infrared NEOWISE as diamonds in the middle panel, and optical difference imaging as pentagons ATLAS in the bottom panel. Marker colours denote the observed photometric wavebands: violet, $v$-band; green, $g$-band; red, $r$-band; coral, $W1$-band; and indigo, $W2$-band; black, $c$-band; magenta, $o$-band. Hollow data outlines represent the time-averaged measurements from ASAS-SN showing 20-day average brightness and daily averages from CRTS and ATLAS.}
    \label{fig:photometry}
\end{figure*}

\subsection{Photometry} \label{photometry}

We gather photometric data for ZTF18abuamgo from several sky surveys covering optical and infrared wavelengths from MJD $= 53627$ (14--09--2005) to MJD $= 60951$ (03--10--2025). Optical time-domain observations were obtained from the Zwicky Transient Facility (ZTF; $r$ and $g$ bands; MJD 58437–60951; \citealt{bellm_zwicky_2019}), which provides high-cadence coverage of the northern sky, and from the All-Sky Automated Survey for SuperNovae (ASAS-SN; $V$ and $g$ bands; MJD 55937–60922; \citealt{shappee_all_2014, kochanek_all-sky_2017}), a global network optimized for bright transient detection. Long-term optical variability was further supplemented with data from the Catalina Real-Time Transient Survey (CRTS; $V$ band; MJD 53627–56591; \citealt{drake_first_2009}). Optical difference imaging is obtained from the Asteroid Terrestrial-impact Last Alert System (ATLAS; $c$- and $o$-bands; MJD 57229-60520; \citealt{tonry_atlas_2018}). The difference magnitude isolates the AGN photometric variability from the constant host galaxy emission. Infrared photometry in the WISE/NEOWISE $W1$ and $W2$ bands (3.4 and 4.6 $\mu$m; MJD 56627–60523; \citealt{wright_wide-field_2010, mainzer_initial_2014}) extends the temporal coverage into the mid-infrared. These surveys collectively enable the multi-wavelength variability of ZTF18abuamgo to be traced over 20 years, as shown in Figure \ref{fig:photometry}. It should be noted that the light curves are not inter-calibrated.

\begin{table*}
    \centering
    \begin{tabular}{lll}
    \hline
    \vtop{\hbox{\strut Modified Julian Date}\hbox{\strut (MJD)}} &
    \vtop{\hbox{\strut Observation Date}\hbox{\strut (DD/MM/YYYY)}} &
    Instrument \\ \hline
    53355 & 16/12/2004 & UK Schmidt Telescope (Six-degree Field Galaxy Survey (6dF)) \\ \hline
    59831 & 09/09/2022 & Nordic Optical Telescope (NOT) \\ \hline
    60550 & 28/08/2024 & ANU 2.3 metre telescope Wide-Field Spectrograph (WiFeS)  \\ \hline
    \end{tabular}
    \caption{Spectroscopic observation dates and instruments used to obtain the spectra shown in Figure~\ref{fig:spectra}.}
    \label{obs_table}
\end{table*}

\begin{figure*}
	\includegraphics[width=15cm]{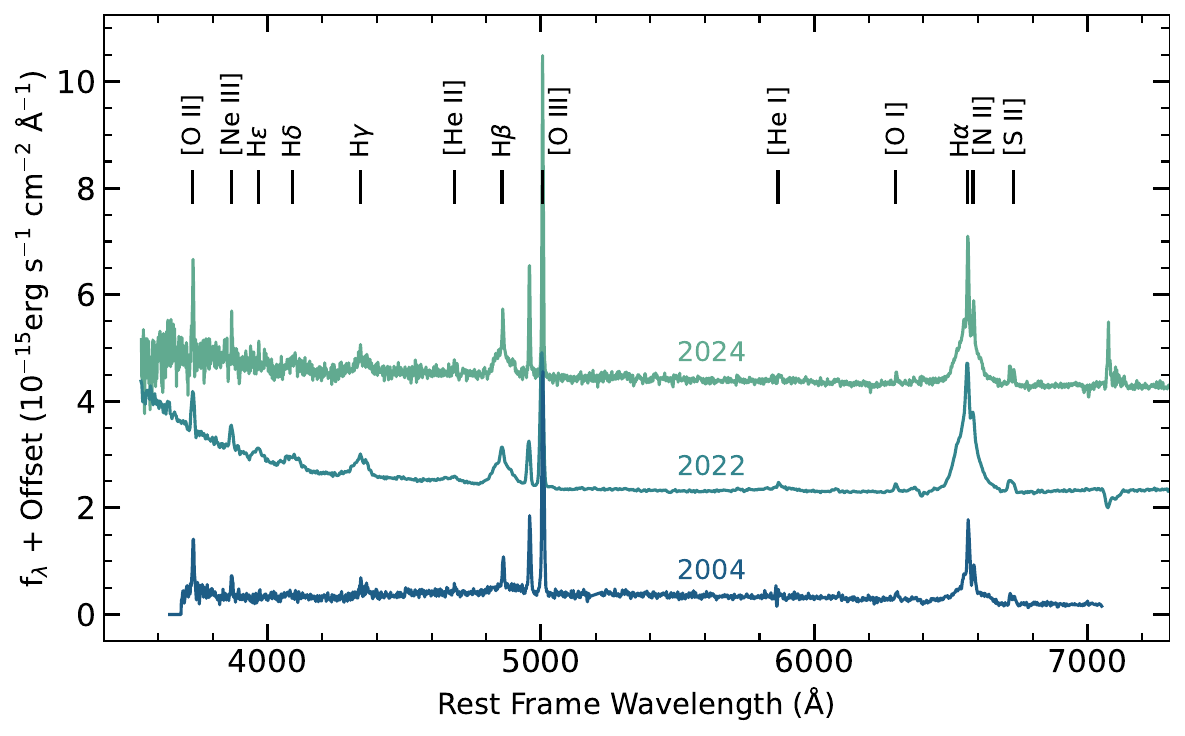}
    \caption{Spectroscopic observations of ZTF18abuamgo from the instruments described in Table \ref{obs_table}. We plot flux density against wavelength for three different epochs: 2004, 2022, and 2024. The most recent, bright-state spectra from 2022, and 2024 are offset by $+1.8$, and $+3.8$ ($10^{-15} \text{erg s}^{-1}\text{cm}^{-1}$\AA), respectively. Narrow and broad emission features are labelled above.}
    \label{fig:spectra}
\end{figure*}

Figure \ref{fig:photometry} displays the apparent magnitude as a function of Modified Julian Date (MJD) and calendar year. The top panel shows optical light curves, while the middle panel shows mid-infrared (MIR) light curves. To reduce scatter in the optical data, we time-average the CRTS and ASAS-SN measurements by binning them into daily and 20-day intervals, respectively; these averaged points are plotted as hollow circle and triangle markers in the top panel.

For ZTF18abuamgo, the combined optical monitoring from CRTS, ASAS-SN, and ZTF spans from 2005 to 2025. The CRTS and ASAS-SN $V$-band light curves exhibit little variability for more than a decade (2005–2018), with total variations smaller than those seen in the other optical bands. However, at the end of CRTS coverage in 2013, we detected a sudden increase in brightness of $0.2$–$0.3$ mag (marked with violet circles). At the start of ZTF observations in late 2018, a short-lived dip followed by a rise of $\approx 0.3$ mag appeared in both $g$- and $r$- bands. These bands then brighten steadily over the next 4 years, until late 2022. During the subsequent bright state, post-2022, the variability amplitude in the ZTF light curves increases relative to the earlier rising phase, reaching nearly 0.5 magnitudes within each observing window. However, because ZTF observations cover only about half of each year, the short-timescale variability before and during the bright state is not well constrained, and the ATLAS data, with comparable temporal gaps and larger photometric uncertainties, do not improve this constraint. The ATLAS difference data, in the bottom panel of Figure \ref{fig:photometry}, provides additional context for the AGN's behaviour pre-ZTF, indicating variability between 2015 and 2018, though with a smaller overall amplitude, of less than 3 magnitudes, than the brightening seen after 2018, which reaches $\sim3.5$ magnitudes. This suggests that the AGN exhibited year-scale variability before the pronounced brightening captured by ZTF.

The middle panel of Figure \ref{fig:photometry} shows MIR light curves from the NEOWISE $W1$- and $W2$-bands ($3.4 \mu$m and $4.6 \mu$m) starting in 2014. These data reveal brightness changes of $\approx0.4$-$0.5$ magnitudes over the monitoring period, with alternating brightening and fading phases. The MIR variations generally track the optical variability seen in the ASAS-SN $g$-, $r$-bands and in ZTF, although the ASAS-SN $V$-band light curve is too noisy to reveal a clear correlation with the MIR trends.

\begin{figure*}
	\includegraphics[width=15cm]{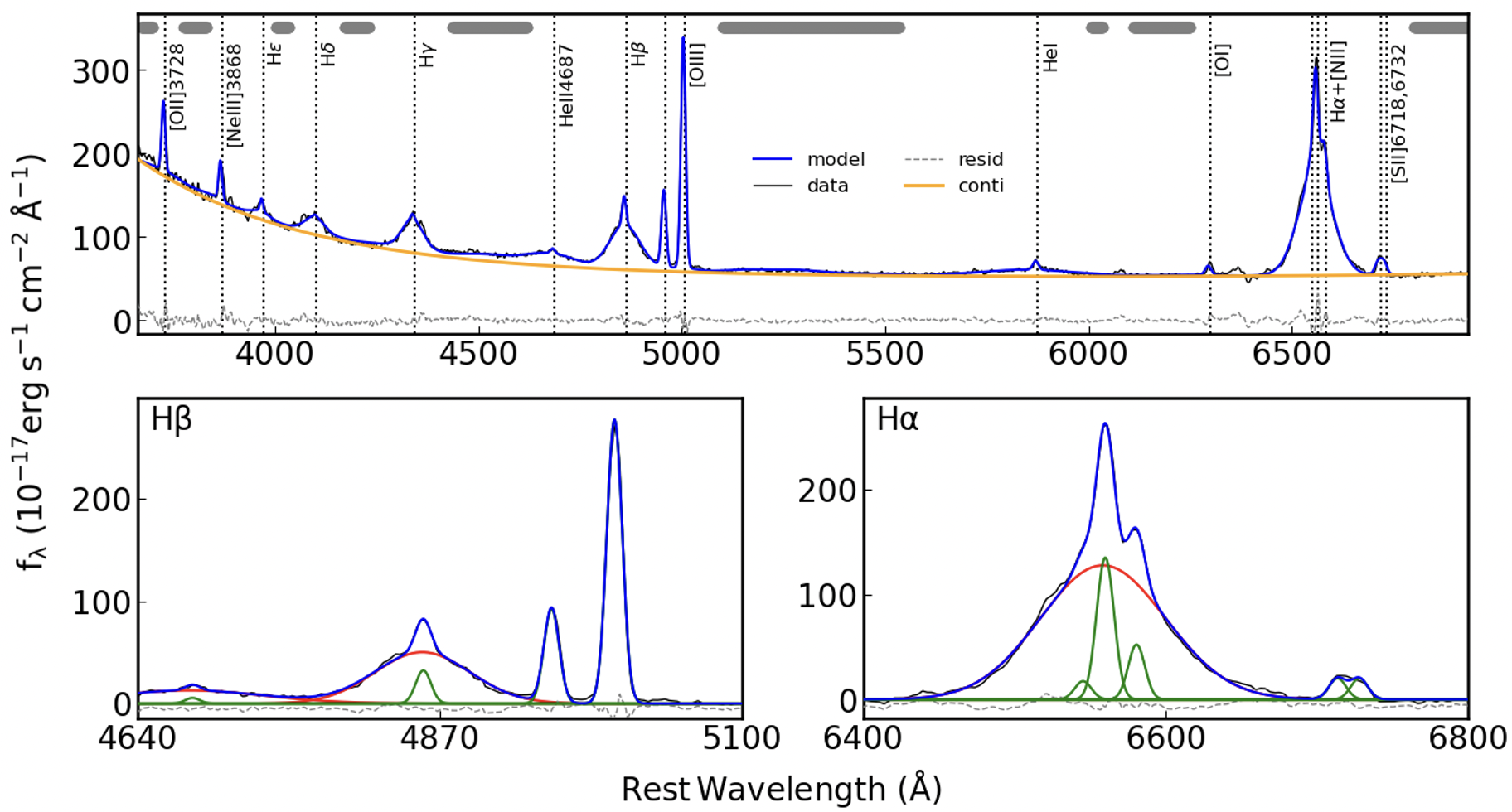}
    \caption{PyQSOFit spectral decomposition of the 2022 optical spectrum. We plot flux density against rest wavelength, with the observed spectrum in black, the best-fitting model in blue, and the residuals as a dotted line. The solid yellow line marks the continuum component of the fit, combined with Gaussian emission lines to reproduce the full spectrum. The lower panels show zoomed-in views of the H$\beta$ and H$\alpha$ line complexes, where the Gaussian components are separated into narrow lines in green and broad lines in red.}
    \label{fig:decomp}
\end{figure*}

\subsection{Spectroscopy} \label{spectroscopy}

To trace the spectral evolution of ZTF18abuamgo, we collect archival data from 2004 and compare this to follow-up spectra taken after the optical brightening event. These spectra are listed in Table \ref{obs_table} and plotted in Figure \ref{fig:spectra}. We use three spectra during our analysis from the Six-degree Field Galaxy Survey (6dF), the Nordic Optical Telescope (NOT), and the Wide-Field Spectrograph (WiFeS) instrument mounted on the Australian National University 2.3 metre (ANU 2.3m) telescope.

The earliest spectroscopic data is provided by 6dF in the Final Redshift Release (DR3) \citep{jones_6df_2004, jones_6df_2009}. In total, 136,304 spectra were observed in the Southern hemisphere between 2001 and 2006 using the UK Schmidt Telescope and the Six-degree Field multi-object fibre spectrograph, with ZTF18abuamgo being observed in December 2004. 6df Observes two wavelength ranges of $\lambda\lambda3900$-$5600\AA$ and $\lambda\lambda5400$-$5600\AA$ which are spliced together for redshift determination. Since the goal of this survey was not to obtain astrophysical measurements, the spectra are not flux-calibrated and can suffer from inconsistent wavelength and flux calibrations between wavelength regimes (see \citealt{hon_broad-line_2024}). \S\ref{calibrate} describes the flux calibration process to obtain measurements from the early archival spectrum.

We obtained a spectrum of ZTF18abuamgo using the Alhambra Faint Object Spectrograph and Camera (ALFOSC) mounted on the NOT. The first follow-up spectrum obtained on 9 September 2022 (MJD 59831) was taken using a 1'' slit mounted with Grism 4 ($\lambda\lambda$3200-9600\AA). The spectrum was bias subtracted, flat-fielded, wavelength calibrated and then extracted using standard routines within {\texttt{IRAF}}. Nightly spectroscopic standard stars observed under the same setup were used for flux calibration.

We obtained a spectrum of ZTF18abuamgo on the 28th of August 2024 utilising the ANU 2.3m WiFeS telescope located at Siding Springs Observatory (SSO) \citep{dopita_wide_2007, dopita_wide_2010, price_converting_2024}. This spectrum was taken in ``Nod \& Shuffle" mode, which results in simultaneous science and sky spectra, of which the sky contribution is subjected during reduction \citep[see Section 2.2 of][for more details]{carr_wifes_2024}. We utilised the $3000$ grating to cover the full 3200-9800 \AA wavelength range. The observations were reduced using the default WiFeS reduction pipeline pyWiFeS to produce calibrated, 3D data cubes and a 2D spectrum was extracted using a region with a size to the seeing on the night \citep{childress_pywifes_2014}. A spectroscopic standard star was also used on the night to calibrate the data.

\subsubsection{Spectral Calibration} \label{calibrate}

To make reliable comparisons of the spectral energy distribution across epochs, we scale the spectra according to their integrated [\ion{O}{iii}]$\lambda5007$ line flux, assuming narrow emission lines are invariant over the 20 years between the first and last observation. To determine which epoch is best calibrated, we convolve the 2022 and 2024 spectra with the ZTF $r$- and $g$-band transmission curves from the SVO Filter Profile Service \citep{rodrigo_svo_2012, rodrigo_svo_2020, rodrigo_photometric_2024} and integrate the result to find the flux, $f$, as measured by ZTF. We convert this to an apparent magnitude with $m = -2.5\text{log}(f / f_{\text{zp}})$ using the zero point flux, $f_{\text{zp}}$, for each filter (also from SVO). Comparing the calculated magnitudes to the photometric ZTF data (Figure \ref{fig:photometry}) collected over $r$- and $g$-bands, the closest match is provided by the 2024 WiFeS spectrum, so we use this as our verified absolute flux calibrated spectrum, and scale the narrow emission lines to match this epoch.

\subsubsection{Spectral Decomposition} \label{spectral_decomp}
To investigate the physical conditions and kinematics within the central region of the AGN, we perform spectral fitting with PyQSOFit to decompose the observed spectra into their constituent components: emission lines, AGN continuum, and host galaxy contribution (see Figure \ref{fig:decomp}). PyQSOFit models the optical spectra of AGN using a chi-squared fitting method \citep{guo_pyqsofit_2018, shen_sloan_2019, ren_prior-informed_2024}. The model includes an AGN continuum, represented by a power law and an \ion{Fe}{ii} template, along with Gaussian components for emission lines. To model the host galaxy, we apply prior-informed fitting based on principal component analysis (PCA) templates \citep{yip_distributions_2004, yip_spectral_2004}. This approach reduces AGN–host redundancy and prevents overfitting by restricting the host component to combinations of galaxy spectra derived from large surveys like SDSS.

To obtain the uncertainties in the line parameters, we use a Monte Carlo method. We generate 200 mock spectra for each epoch with Gaussian noise applied to the original spectrum model generated with PyQSOFit. We fit single Gaussian components to emission lines, including broad emission, which, based on the fit residuals, do not require more than one component to account for skewness, which is not the case for all AGN (see \citealt{wamsteker_ultraviolet_1990, korista_broad_1992}). Narrow line widths are tied to one another, with separate widths allowed for the $[\ion{O}{iii}]\lambda5007$ and $[\ion{O}{iii}]\lambda4959$ doublet, due to their significantly stronger flux making fitting difficult. The following doublets have their flux ratios fixed at: $[\ion{O}{iii}]\lambda5007 / [\ion{O}{iii}]\lambda4959 = 3$; $[\ion{N}{ii}]\lambda6583 / [\ion{N}{ii}]\lambda6548 = 3$; and $[\ion{S}{ii}]\lambda6716 / [\ion{S}{ii}]\lambda6731 = 1$ (similar to fits by \citealt{belli_star_2024}). The [\ion{S}{ii]} doublet can take a range of values, however in our case with low signal-to-noise and blended profiles (see Figure \ref{fig:spectra} and Figure \ref{fig:decomp}), we approximate them to be 1:1. We allow the width of broad H$\alpha$ to remain independent of other broad Balmer lines, which are tied in width, since H$\alpha$ is significantly more luminous and has been observed to exhibit slightly different widths in AGN, as reported in the literature \citep{greene_estimating_2005-1}.

\section{Results} \label{results}

\subsection{Changing-look Transition} \label{changing_look}

\subsubsection{Spectral Transformation}\label{Spectral transformation}

We use our spectroscopic observations to compare the transformation observed in our target to other CLAGN in the literature to: (1) probe the driving mechanism behind the transient; (2) to extract properties of the BLR; and (3) to estimate the mass of the central SMBH. Figure \ref{fig:spectra} shows our spectra for the three observations, listed in Table \ref{obs_table}: 6dF (2004), NOT (2022), and WiFeS (2024). The archival 2004 spectrum catches the AGN in its dim state, with no detectable broad Balmer emission beyond faint H$\alpha$ and H$\beta$. In 2022, the spectrum changes; we can see the appearance of various broad emission lines and a change in the shape of the continuum. The emergence of broad Balmer emission from H$\alpha$ (electron transition: $n=3\rightarrow2$) to H$\epsilon$ ($n=7\rightarrow2$) indicates a possible changing-look transition to a bright state and is accompanied by an optically blue continuum tail visible from accretion disc-driven UV emission \citep{czerny_role_2006}. This power-law tail feature is not so apparent 2 years later, in the 2024 spectrum, while the broad emission lines persist, maintaining their bright state. The AGN continuum power law component is defined in PyQSOFit as: $f_{\lambda} = A(\lambda/\lambda_0)^{(\alpha)}$, where $\lambda_0 = 3000$\AA, and the normalisation parameters, $A$, and slope, $\alpha$ are values to be found. Between 2022 and 2024, the normalisation changes from $A_{2022} = 870\pm60$ to $A_{2024} = 800\pm100$, and slope, $\alpha_{2022}=-8.2\pm0.4$ to $\alpha_{2024}=-12.1\pm0.6$.

\begin{figure}
	\includegraphics[width=\columnwidth]{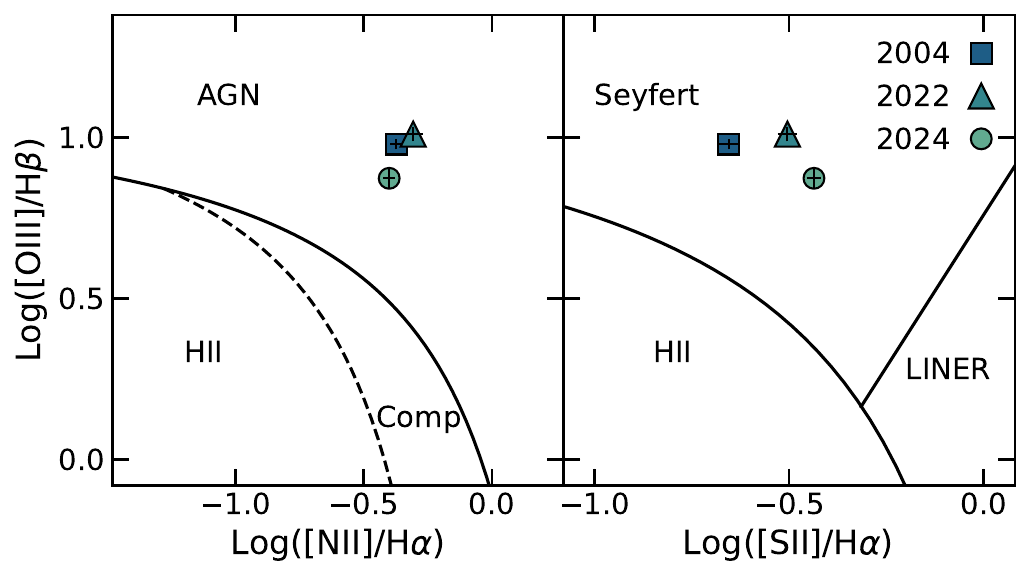}
    \caption{The BPT diagram identifies the dominant excitation mechanism of narrow-line emission using emission-line flux ratios. Due to the location of narrow emission from the central engine, these line ratios probe the emission mechanism over thousands of years. We adopt the solid and dashed diagnostic boundaries from \citealt{kewley_host_2006} and plot the emission-line ratios from our 2004, 2022, and 2024 spectra as squares, triangles, and circles, respectively. In all epochs, the line ratios fall in the Seyfert-AGN region, indicating the presence of AGN activity before the changing-look transition.}
    \label{fig:bpt}
\end{figure}

We verify the nucleus was active in ZTF18abuamgo before the apparent state change using a Baldwin, Phillips and Terlevich plot (BPT; \citet{baldwin_classification_1981}) in Figure \ref{fig:bpt}. This diagram is used to identify the dominant excitation mechanism of optical narrow emission lines. The narrow lines are emitted further from the central engine than broad lines, and therefore trace longer activity periods of order $10^5$ years \citep{schawinski_active_2015}. From the BPT diagram, we determine that the main excitation mechanism is a Seyfert-like AGN emission. Therefore, the observed continuum brightening and broad emission transformations in ZTF18abuamgo are indicative of an intrinsic change to the pre-existing AGN rather than an awakening of a dormant black hole.

To classify the spectral state of each epoch, and to contextualise the transformation based on line variability, we use the flux ratio of [\ion{O}{iii}]$\lambda5007$ to H$\beta$, established by \citet{osterbrock_spectrophotometry_1977, osterbrock_seyfert_1981-1}, and follow \citet{whittle_virial_1992}'s convention for AGN classification types. The dim-state 2004 spectrum can be classified as a Seyfert Type 1.5, and this changes to Type 1.2 for the bright-state spectra in 2022 and 2024 after the CL transition. This quantifies the shift from an AGN with comparatively weaker broad-line emission to one dominated by strong broad lines, as the broad H$\alpha$ flux increases by more than a factor of three.

\subsubsection{Changing-look or Extreme Variability?} \label{CL_or_EVQ}

When selecting for CLAGN in large AGN samples, studies typically quantify the variability in broad line flux by statistical significance of the amplitude change (e.g. see examples from \citealt{macleod_changing-look_2019, green_time_2022, yang_galaxies_2025}). ZTF18abuamgo was identified as a candidate nuclear transient based on its light curve and then followed up spectroscopically to confirm its CLAGN nature. Here, we test whether this nuclear transient would have been selected as a CLAGN from large surveys. This is particularly interesting for this source due to the comparatively small spectral transition between the dim and bright states: from a Type 1.5 in 2004 to a Type 1.2 AGN in 2022.

The approach taken in selecting CLAGN through line flux variability is not always the same between studies. \citet{zeltyn_exploring_2024} identify CLAGN from SDSS V by defining the quality $C(\text{line}) \equiv F_{\text{bright}}/F_{\text{dim}} - \Delta(F_{\text{bright}}/F_{\text{dim}})$ and select candidates with $C(\text{line}) > 2$. This equation takes the integrated line flux from the same broad emission line in spectra over two epochs, $F_{\text{bright, dim}}$. The second term takes the $1\sigma$ uncertainty on their ratio, $\Delta(F_{\text{bright}}/F_{\text{dim}})$. In the case of ZTF18abuamgo, we measure, between 2004 and 2022 spectra, $C(\text{H}\alpha) = 3.8$ and $C(\text{H}\beta) = 2.4$. This places ZTF18abuamgo in the category of CLAGN as opposed to a highly variable AGN, or EVQ, based on the H$\alpha$ and H$\beta$ line variability. Nonetheless, both EVQs and CLAGN challenge the understanding of AGN variability and may therefore offer insights into the underlying mechanisms of accretion disc emission in AGN.

\subsubsection{Changing-look Timescale}

The changing-look timescale refers to the period over which an AGN undergoes significant spectral transitions, such as the appearance or disappearance of broad emission lines and changes in continuum brightness. It is typically defined as the time between observations that captures a transformation, for example, from a Type 1 to a Type 2 spectrum (e.g. \citealt{yang_discovery_2018, jana_investigating_2024}). Capturing this transition requires sufficiently frequent spectroscopic monitoring, which is often serendipitous. The timescale of ZTF18abuamgo's transition, based on spectroscopic analysis, is 18 years between 2004 and 2022, as confirmed by the broad emission line fitting. 

We are limited here due to the archival spectra being decades old, so we use photometric data to place tighter constraints on the changing-look timescale. Using the multi-wavelength light curves between spectroscopic observations, presented in Figure \ref{fig:photometry}, we estimate when the changing-look transition most likely occurred. The earliest available photometry, from CRTS, begins several months after the first spectroscopic observation in 2004, when the AGN was in a dim state. For more than a decade following this, we see no evidence for large-amplitude variability. The most significant change occurs beginning in 2018, when the optical brightness increases by approximately 1 magnitude in the ZTF $g$-band. This is accompanied by brightening in ZTF $r$-band; ASAS-SN's $g$-band; NEOWISE's infrared $W1$ and $W2$-bands; and ATLAS $o$ and $c$-bands difference imaging. This brightening phase reaches a maximum in 2022, when the follow-up 2022 spectra confirms a changing-look transition has occurred. We therefore estimate the changing-look to have occurred between 2018-2022, suggesting a significantly shorter timescale for changing-look of 4 years, compared to the conservative 18-year estimate.

\subsubsection{Accretion Rate Change}

We track the evolution of the accretion rate to constrain possible changing-look mechanisms by calculating the Eddington ratio at all three epochs in 2004, 2022, and 2024. We take continuum measurements at 5100\,\AA\, because of its proximity to the calibrated [\ion{O}{iii}]$\lambda5007$ narrow emission line (see Section \S\ref{calibrate}). The 2024 WiFeS bright-state spectrum provides the opportunity to measure the host-contribution with less dominant AGN features, such as the blue-tail from a bright UV-disc, which is less prominent than in 2022. This decreases the degeneracy between host and AGN continuum contributions, following the same principle as \citet{runnoe_now_2016}, who take their host measurements exclusively from dim-state spectra due to the lower AGN flux contribution. Using the prior-informed host decomposition of the 2024 spectrum with PyQSOFit, we find the host luminosity to be $L_{5100,\text{Host}}=1.56\times10^{43} \text{erg s}^{-1}$. The optical continuum luminosity from the AGN in the three epochs is then found by subtracting the host contribution: $L_{5100,\text{AGN}} = (7.58\pm1.12)\times10^{42}  \text{ergs s}^{-1}$, in 2004; increasing to $(2.38\pm0.11)\times10^{43}$ and $(2.44\pm0.17)\times10^{43} \text{ergs s}^{-1}$ in 2022 and 2024 respectively. 

To calculate the Eddington ratio, we first determine the bolometric luminosity, $L_{\text{bol}}$, which is the AGN luminosity emitted over all wavelengths. We approximate this using \citet{netzer_bolometric_2019}'s $L_{5100}$ to $L_{\text{bol}}$ conversion, assuming an optically thick, geometrically thin accretion disc and time-independent accretion rate: $L_{\text{bol}} = k_{\text{bol}} \times L_{5100}$. At each of our three spectral epochs, the optical continuum, $L_{5100}$, is multiplied by a bolometric correction factor, $k_{\text{bol}}$, which takes the value of
\begin{equation}
    k_{\text{bol}} = 40\times \big( L_{5100}/10^{42} \,\text{erg sec}^{-1} \big)^{-0.2}.
\end{equation}

We measure the bolometric luminosity to be $(2.0\pm0.2)\times10^{44}$, $(5.0\pm0.2)\times10^{44}$ and $(5.2\pm0.3)\times10^{44}$erg s$^{-1}$ in the dim (2004) and bright state (2022, 2024) spectra, respectively. It should be noted that the assumption of an optically thick, geometrically thin accretion disc may conflict with current concepts of CLAGN mechanisms involving different disc dynamics, as well as additional uncertainties from the unknown black hole mass, spin, and inclination. Predicting the bolometric luminosity, for example, is dependent on the shape of the spectral energy distribution, which is known to change under CLAGN transitions (e.g. \citealt{runnoe_updating_2012}). This is unavoidable; however, as many effects are systematic between epochs of the same source, so approximate comparisons can still be made. We calculate the Eddington ratio:
\begin{equation}
    \lambda_{\text{Edd}} = L_{\text{bol}}/1.26 \times 10^{38}M_{\text{BH}}
    \label{EddingtonRatio}
\end{equation}
using the bolometric luminosity, $L_{\text{bol}}$ derived above, and the black hole mass, $M_{\text{BH}} = (5.04\pm0.38)\times10^7 M_{\odot}$, calculated in Section \S\ref{bh_mass}. In the dim-state $\lambda_{\text{Edd},2004} = 0.032\pm0.005$ which increases to $\lambda_{\text{Edd},2022} = 0.080\pm0.007$ and $\lambda_{\text{Edd},2024} = 0.081\pm0.010$ in the bright states. Strict upper and lower limits on these values can be found with the basic assumption that the host remains constant over all epochs. Calculating the Eddington ratio with a dominant AGN contribution (i.e. no host accounted for; $L_{5100,\text{AGN}} = L_{5100}$) gives upper limits (ul) on the Eddington ratio of $\lambda_{\text{Edd,ul}} = 0.078$, $0.12$, $0.12$ in 2004, 2022 and 2024 respectively. On the other hand, the brightest host galaxy contribution possible is equal to the lowest continuum measured, from the dim-state 2004 spectrum. This gives lower limits (ll) on Eddington ratio: $\lambda_{\text{Edd,ll}} = 0.0$, $0.041$, $0.044$ in 2004, 2022, and 2024, respectively.

\subsection{Broad Line Region Temperature} \label{blr}

We extract properties of the BLR during the two bright-state epochs of ZTF18abuamgo to investigate how its variable BLR compares to those in typical AGN that do not exhibit changing-look behaviour. Specifically, we follow the approach of \citet{popovic_balmer_2003}, applying the Boltzmann plot (BP) method, previously used to analyse the broad Balmer line series in standard AGN \citep{popovic_balmer_2003, popovic_broad_2006, mura_detailed_2007, ilic_analysis_2012}, and apply it here for the first time to a CLAGN.

The Boltzmann plot method relates the normalised line intensity, $I_n$, plotted in logarithmic space, to the upper energy level of the corresponding transition, $E_u$. Under the assumption that the BLR plasma is optically thin and in partial local thermodynamic equilibrium (PLTE), such that collisional processes dominate the population of excited states, the level populations follow a Boltzmann distribution. In this case, the Balmer line intensities satisfy a linear relation between $\log I_n$ and $E_u$, allowing the electron temperature to be inferred from the slope of the best-fitting line \citep{popovic_broad_2006, mura_detailed_2007}:
\begin{equation}
    \text{log}(I_n) = \text{log} \bigg( \frac{F_{ul}\times\lambda}{g_uA_{ul}} \bigg) = B - A E_u
    \label{intensity}
\end{equation}

where $F_{ul}$ is the measured flux of the transition from upper to lower level, and $\lambda$, $g_u$, and $A_{ul}$ are the transition wavelength, upper-level statistical weight, and transition probability, respectively. The constants $A$ and $B$ are determined from a linear regression fit to the Boltzmann plot, with $A$ known as the temperature parameter.

This method further assumes that all Balmer emission lines originate in the same BLR region \citep{mura_optical_2011}, although in practice different transitions may arise at slightly different radii or under different physical conditions. In our spectral fits with PyQSOFit, the broad line widths are tied, apart from H$\alpha$, which is fit independently due to its much higher luminosity.

\begin{figure*}
	\includegraphics[width=15cm]{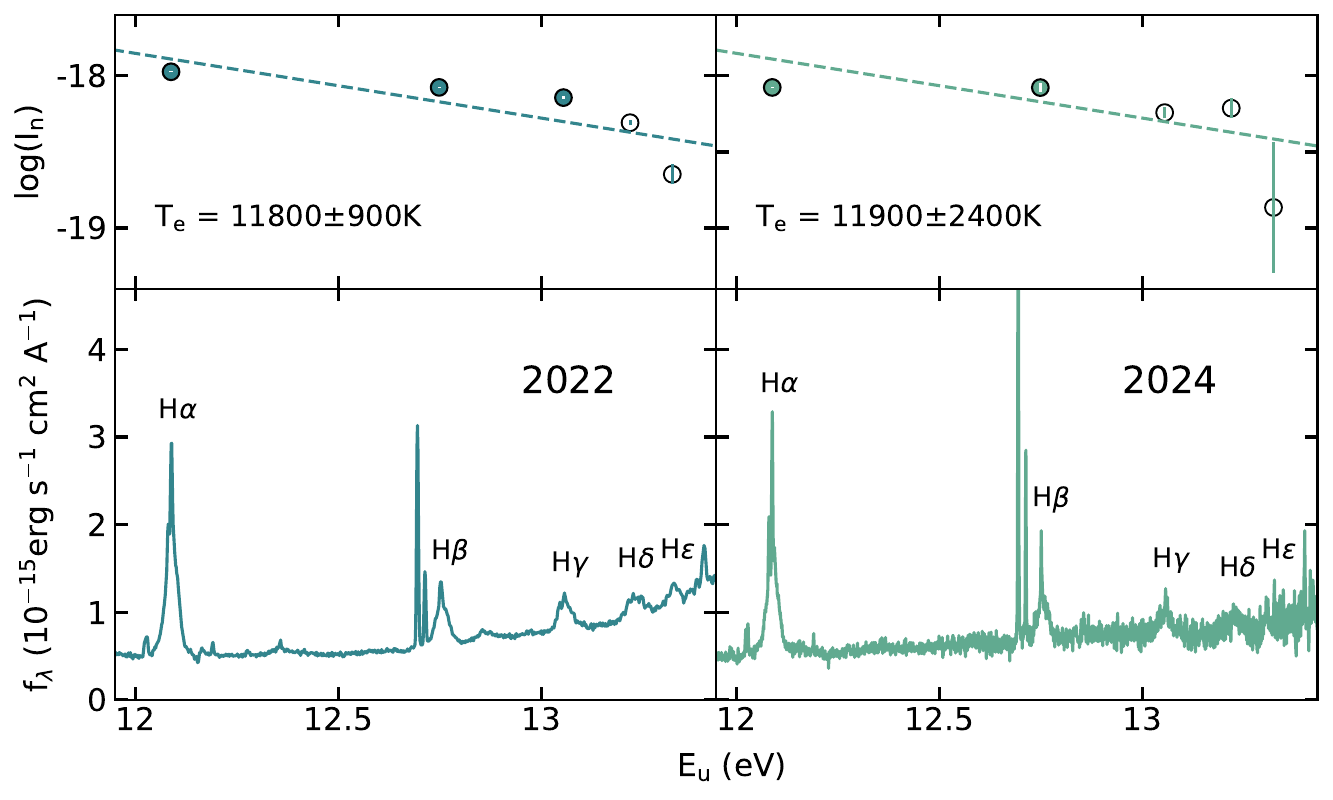}
    \caption{Boltzmann plot for 2022 (left) and 2024 (right) with normalised line intensity plotted against upper energy level of the Hydrogen Balmer transitions (top) and the Balmer lines present in their respective spectra using flux density over the transition energy (bottom). Solid data points have a signal-to-noise ratio greater than 3, whereas white points have a ratio below 3. The bottom plots are simply representatives of the Balmer lines in their spectra; only discrete upper energy states are physically motivated. The electron temperatures of each epoch, as calculated from the linear regression gradient, T$_{\text{e}}$, are stated in the top panels for both Boltzmann plots.}
    \label{fig:bp}
\end{figure*}

We show the Boltzmann plots generated with the Balmer emission from our bright state 2022 (left) and 2024 (right) spectra in the top panels of Figure \ref{fig:bp}. The corresponding spectral energy distributions are plotted in the bottom panels, where we recast the x-axis as the photon transition energy levels, which have no physical meaning apart from the discrete energy transitions labelled. The circular data points in the top panels represent the log line intensity for each broad Balmer line (H$\alpha$ ($n=3\rightarrow2$) to H$\epsilon$ ($n=7\rightarrow2$)), plotted against the corresponding upper energy level. We denote lines with a higher signal-to-noise ratio of at least $3$ to be solid black points, whilst lines with a signal-to-noise ratio below $3$ are not filled in. The uncertainty in normalised line intensity is plotted over the data points in the Boltzmann plots in white and black. In the spectral energy distributions, we see a decrease in signal-to-noise of the emission lines at higher photon energies (lower wavelengths). This is due to both the increasing continuum noise in that regime and the decreasing flux of the higher-order Balmer emission lines. The dim state, 2004 spectrum cannot be used for this analysis due to the lack of Balmer emission; only the H$\alpha$ and faint H$\beta$, are detected. 

Deviations from the optically thin and PLTE assumptions can produce the curvature we see in Figure \ref{fig:bp}, leading to unreliable temperature estimates \citep{mura_optical_2011}. In particular, higher-order Balmer lines are often weak and more strongly affected by measurement uncertainties, reducing the robustness of the linear fit. Therefore, the Boltzmann plot-derived BLR temperatures should be interpreted as approximate, comparative indicators rather than precise physical measurements.

To estimate the BLR temperature, the temperature parameter, $A$, from Equation \ref{intensity}, is measured from the Boltzmann plot gradient, shown in the dashed straight lines fitted in both epochs in Figure \ref{fig:bp}. We measure the temperature parameter to be $ A\sim 0.4$ in both cases ($A=0.43\pm0.03$, $A=0.42\pm0.08$). The electron temperatures are then calculated as $T_{\text{e}} = \text{log}(e)/(kA)$ with the Boltzmann constant, $k$ \citep{popovic_balmer_2003}, to be $11,838\pm928$K and $11,903\pm2411$K in 2022 and 2024, falling within the range of $7,000$-$37,000$K for AGN reported by \citet{ilic_analysis_2012}. The temperatures measured in 2022 and 2024 are consistent within the uncertainties, showing the temperature of the BLR remains constant even with a highly variable continuum in the bright state.

\subsection{Single-epoch Black Hole Mass Estimates} \label{bh_mass}

To test the reliability of single-epoch SMBH mass measurements, we analyse the spectra of ZTF18abuamgo obtained in 2004, 2022, and 2024. This allows us to estimate black hole masses at different epochs and evaluate the efficacy of using highly variable systems such as CLAGN for mass determinations. Our approach follows several well-established correlations between black hole mass and broad emission line widths and luminosities \citep{greene_estimating_2005-1, bonta_estimating_2024, lamassa_stripe_2024}.

These estimates assume a virialised BLR, supported by reverberation mapping results showing that BLR size scales with luminosity and that broad-line widths trace the virial velocity \citep{peterson_central_2004, bentz_low-luminosity_2013}. In this framework, the width of H$\alpha$ provides a proxy for the virial velocity, while line or continuum luminosities (e.g. $L_{5100}$) give an estimate of the BLR radius through the $R$–$L$ relation \citep{kaspi_reverberation_2000,bentz_low-luminosity_2013}. As is standard, such analysis is typically carried out in bright states of CLAGN, when broad emission lines are strong enough for reliable measurements, rather than in dim Type-1.8/1.9/2 states, where the reduced flux makes such measurements uncertain or impossible. However, ZTF18abuamgo, which retains Type-1.5 characteristics even in its dim state, offers a rare opportunity to investigate the impact of deriving black hole masses from both dim and bright states using the broad H$\alpha$ Balmer line, without a severe loss in signal-to-noise.

The mass estimates are grounded in the same virial framework that links BLR size and velocity dispersion to black hole mass, which can be written in the general form:
\begin{equation}
    \text{Log} \bigg( \frac{M_{\text{BH}}}{M_{\odot}} \bigg) = a + \text{Log}\Bigg[\bigg( \frac{L_{\text{Line}}}{10^{b} \text{ ergs s}^{-1}} \bigg)^c \bigg( \frac{\text{FWHM}_{\text{Line}}}{10^{d} \text{ km s}^{-1}} \bigg)^e\Bigg].
    \label{mass}
\end{equation}

\begin{figure}
	\includegraphics[width=\columnwidth]{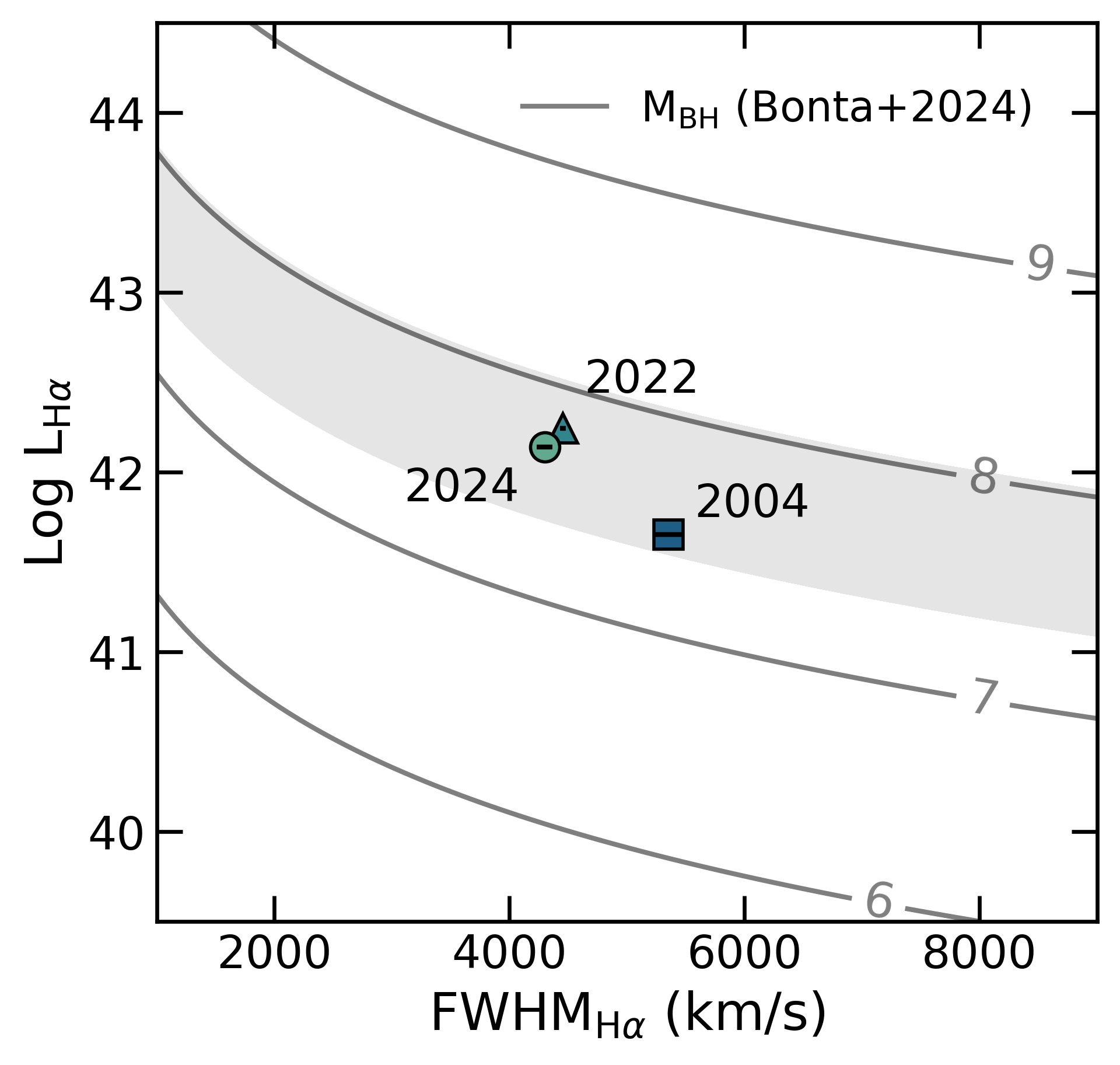}
    \centering
    \caption{Luminosity of the broad H$\alpha$ emission line (in log space) as a function of its FWHM, with contours showing the logarithm of the black hole mass (units of log$M_{\text{BH}}/M_{\odot}$). The three data points indicate the H$\alpha$ measured in the three spectra: dim-state 2004 (square), bright-state 2022 (triangle), and 2024 (circle). The intrinsic scatter associated with the mass measurement from \citealt{bonta_estimating_2024}'s study of $0.332$ dex is represented by the shaded region centred on the mass we predict with H$\alpha$ in 2024.}
    \label{fig:mass}
\end{figure}

Different studies use slightly different calibrations for the single-epoch black hole mass formula (Equation \ref{mass}), resulting in varying constants $a$, $b$, $c$, $d$, and $e$. These differences arise from fitting AGN samples spanning a range of redshifts, luminosities, and observational conditions. In most cases, $L_{\text{Line}}$ and $\text{FWHM}_{\text{Line}}$ represent the luminosity (in $\text{erg s}^{-1}$) and full width at half maximum (FWHM) (in $\text{km s}^{-1}$), respectively, of a broad emission line, such as H$\beta$. For \citet{lamassa_stripe_2024}, $L_{\text{Line}}$ may be given as the continuum luminosity, $L_{5100}$ instead. To test whether consistent black hole masses are recovered across different AGN states in changing-look sources, we calculate single-epoch mass estimates for ZTF18abuamgo using spectra from 2004 (dim), 2022 (bright), and 2024 (bright). Line properties are measured using PyQSOFit (see \S\ref{spectral_decomp}), and applied to several published mass relations. While each relation varies slightly in constants, they are all reformulations of Equation \ref{mass}. For our main comparison (Figure \ref{fig:mass}), we adopt the mass estimate of \citealt{bonta_estimating_2024}, which reports the smallest intrinsic scatter ($0.332$ dex). This scatter dominates the total uncertainty once combined in quadrature with our line luminosity and width measurement errors from the spectra (see \S\ref{spectral_decomp}). As a result, this choice provides the strictest constraints and maximises sensitivity to detecting inconsistencies among mass estimates from different spectra.

Figure \ref{fig:mass} shows the luminosity–FWHM parameter space for the broad H$\alpha$ line, with our measurements for each epoch overplotted. The background contours represent black hole mass predictions (in $\log M_{\text{BH}}/M_\odot$) based on the H$\alpha$ single-epoch relation of \citet{bonta_estimating_2024}. This model was calibrated using reverberation-mapped AGN samples available up to 2019, including the SDSS Reverberation Mapping Project \citep{shen_sloan_2024}. We adopt their final calibration constants ($a = 7.37$, $b = 42$, $c = 0.812$, $d = 3.5$, $e = 1.634$), and include the intrinsic scatter of $0.332$ dex as a shaded region around the 2024 prediction. This scatter is added in quadrature to our line measurement uncertainties.

Between 2004 and 2024, we observe the H$\alpha$ emission becomes both brighter and narrower. The FWHM decreases between the dim and bright states in 2004 and 2022, from $5400\pm100$km s$^{-1}$ to $4450\pm20$km s$^{-1}$, respectively, as it undergoes the spectral transition. Meanwhile, the broad H$\alpha$ luminosity increases by a factor of nearly $3.9$ times from $4.5\times10^{41}$ to $1.7\times10^{42}\text{ergs s}^{-1}$. Despite these changes in individual line properties, the derived black hole masses remain broadly consistent across epochs when uncertainties and intrinsic scatter are considered. The additional models we test produce the same results. The 2004 dim-state spectrum yields the lowest mass estimate, with a black hole mass of $2.90 \times 10^7 M_\odot$. The H$\alpha$ measured in the bright state spectra of 2022 and 2024 give masses of $6.47 \times 10^7 M_\odot$ and $5.04 \times 10^7 M_\odot$. While these differences highlight the sensitivity of single-epoch methods to line variability, they remain within the typical intrinsic scatter of such estimators ($\sim0.4$–$0.5$ dex; \citealt{vestergaard_determining_2006}).

We adopt the 2024 mass estimate in our Eddington ratio calculations (see \S\ref{changing_look}, Equation \ref{EddingtonRatio}), as it is consistent with results from all three epochs and carries the smallest model-dependent uncertainty. Overall, these results demonstrate that ZTF18abuamgo, despite its spectral evolution from a Type 1.5 in 2004 to a brighter Type 1.2 in 2024, yields robust and consistent black hole mass estimates across states when single-epoch methods are applied.

\section{Discussion} \label{discussion}

\subsection{Changing-look Mechanism}

Constraining the physical mechanism responsible for the changing-look transition is very challenging; therefore, we use a combination of spectral and photometric observations to derive clues on the physical mechanisms involved. In a changing-obscuration scenario, where continuum and broad line variability were due to obscuration by dusty torus material, we would expect corresponding photometric signatures in the optical and MIR bands. Dust moving across the line of sight would reprocess UV and optical light into longer wavelengths \citep{lopez-rodriguez_origin_2018}, thereby dimming the optical flux from the disc while simultaneously enhancing infrared re-radiation, producing an anti-correlation between the bands \citep{denney_typecasting_2014}. In contrast, intrinsic changes to the disc emission, in a changing-state event, would drive correlated variability, with both optical and MIR flux increasing together as the torus responds to enhanced disc output. The optical-MIR photometry presented in Figure \ref{fig:photometry} shows correlated variability between wavebands as the broad emission lines brighten by a factor of up to $3.8$ times, for H$\alpha$, and AGN continuum increases from $L_{5100,\text{AGN}} = (7\pm1)\times10^{42}$ to $(2.4\pm0.1)\times10^{43} \text{erg s}^{-1}$. The largest photometric change seen between dim and bright state spectroscopic observations was between 2018 and 2022, both optical and MIR bands brightened by about $\sim1$ magnitude and $\sim0.5$ magnitudes, respectively. These correlated variations are consistent with intrinsic accretion disc variability, so we classify ZTF18abuamgo as a changing-state AGN rather than a changing-obscuration object. Our lower limit for the changing-look timescale of 4-years, based on the highest amplitude photometric variability observed from ZTF (2018-2022), is consistent with the several-year timescale seen by \citet{jana_investigating_2024} from samples of CLAGN with more regular spectroscopic coverage.

The distinct optical blue tail from a UV peak in the 2022 continuum (see Figure \ref{fig:spectra}) is an indicator of brighter accretion disc emission (and ionising continuum) compared to the dim state, possibly driving the broad line emission. However, the blue tail appears suppressed in the 2024 bright-state spectrum, only two years later, demonstrating that the continuum is highly variable in the bright phase, while the broad emission lines are maintained. This disconnect between the accretion disc flux and broad line change is similar to what has been observed in the CLAGN Mrk 590 (\citealt{raimundo_muse_2019}), which has shown variability in the broad lines without significant variability in the accretion disc flux. The flux variability can also be seen in the optical ZTF and ATLAS light curves in Figure \ref{fig:photometry}. Optical variability is observed by ZTF during the 6-month observation period for each year, reaching fluctuations in apparent magnitude of almost $0.5$ magnitudes between 2022 and 2025.

Using the Eddington ratio as a proxy for accretion rate provides a useful way to compare CLAGN across a wide range of black hole masses, since the dimensionless scaling accounts for differences in $M_{\text{BH}}$. From the optical $L_{5100}$ continuum of the three spectra, we estimate $\lambda_{\text{Edd},2004} = 0.032 \pm 0.005$, rising to $\lambda_{\text{Edd},2022} = 0.080 \pm 0.007$ and $\lambda_{\text{Edd},2024} = 0.081 \pm 0.010$. Although ZTF18abuamgo undergoes only a modest transition from Type 1.5 to 1.2, the relative change in Eddington ratio is comparable in magnitude to those reported for CLAGN with more dramatic type changes \citep{macleod_changing-look_2019, green_time_2022, lyu_wise_2022, dong_newly_2025}.

For a direct comparison, \citealt{jana_investigating_2024} studied 20 CLAGN from the BAT AGN Spectroscopic Survey (BASS) that transitioned between Type 1.0 and 1.8/1.9/2.0. The transition Eddington ratio, inferred as the midpoint between the Eddington ratios measured in the bright (Type 1.0) and dim (Type 1.8/1.9/2.0) states, was reported to have a median value of $\sim0.01$. Our inferred transition ratio of $0.056$ is therefore somewhat higher. However, this comparison should be treated with caution, since ZTF18abuamgo exhibits only a modest change from Type 1.5 to 1.2, whereas the BASS sample involves larger transitions from Type 1.0 to 1.8/1.9/2.0. In addition, any estimate of $\lambda_{\text{Edd}}$ carries uncertainties related to bolometric corrections, spectral energy distribution changes, and black hole mass measurements. Within the sample, three objects, NGC 5548, Mrk 1018, and Fairall 9, show transition ratios similar to ours, suggesting that ZTF18abuamgo’s behaviour is nonetheless consistent with the observed diversity of CLAGN accretion-state changes.

Although we cannot make definitive statements about the mechanism driving the changing-look behaviour beyond accretion rate variations, we can rule out some proposed scenarios linked to rapid accretion rate changes. In a disc-wind BLR scenario, outflows from the accretion disc sustain clumpy BLR clouds within the region of gravitational dominance of the central SMBH (e.g. \citet{nicastro_broad_2000, elitzur_agn-obscuring_2006}). This model requires a minimum bolometric luminosity of $L_{\text{crit}} \simeq 1.5 \times 10^{40}  \text{erg s}^{-1}$ for a black hole mass of $M = 5\times 10^7 M_{\odot}$ \citep{elitzur_disappearance_2009}. This threshold is roughly four orders of magnitude below the bolometric luminosity of ZTF18abuamgo, suggesting that a lack of disc outflows is unlikely to be responsible for any BLR cloud diminution. Another scenario that appears unlikely is disc perturbation caused by a tidal disruption event (e.g. \citealt{eracleous_elliptical_1995, merloni_tidal_2015, blanchard_ps16dtm_2017}). Such events, produced by stars passing close to SMBHs with mass $M_{\text{BH}} \lesssim 10^{8}M_{\odot}$, typically generate light curves with a rapid rise to peak brightness in a matter of weeks, followed by a power-law decline (see review of TDEs; \citealt{gezari_tidal_2021}). ZTF18abuamgo shows no evidence for this behaviour in its optical or infrared light curves.

\subsection{Insights to the Broad Line Regions via Boltzmann Plots}
A novel aspect of this study is the application of Boltzmann plot analysis (Figure \ref{fig:bp}) to the broad Balmer emission lines during the bright states, providing direct constraints on the physical conditions within the BLR. ZTF18abuamgo offers a rare opportunity to do this due to the presence of five detectable Balmer emission lines, although the quality of the fit in the Boltzmann plots is relatively poor (see also \citealt{mura_detailed_2007}). In our case, we are limited by the low luminosity of some Balmer lines and increased noise in the continuum at lower wavelengths. This is not a problem unique to ZTF18abuamgo; however, we compare our optical spectroscopic data to a sample of 82 CLAGN from \citealt{yang_galaxies_2025}, to assess how often BP analysis might be applied to other CLAGN. From the visual inspection of those 82 sources, none show five distinguishable broad Balmer lines (up to H$\epsilon$), and only 32 show four lines (up to H$\delta$). We therefore assume that the probability of being able to use CLAGN with optical spectra for this analysis of the BLR, using the BP method with only four points to constrain the temperature, is approximately $40\%$, and with an ideal five detectable broad Balmer lines, is significantly lower, $\lesssim1\%$. The aim of investigating the temperature of the BLR throughout a changing-look transition, using the dim state spectrum as well, only exacerbates the issue because of the lower luminosity of all the Balmer emission lines. 
Attempting to investigate BLR temperatures throughout a full changing-look transition, using the dim state spectrum as well, exacerbates the problem, since the Balmer emission lines are significantly fainter.

We successfully fit a linear trend in the Boltzmann plots between the log-normalised line intensity versus upper energy levels of the Balmer transitions (H$\alpha$ to H$\epsilon$). The presence of a linear regression indicates that PLTE conditions are met, allowing us to estimate the electron temperature \citep{popovic_balmer_2003}. The derived electron temperatures of ZTF18abuamgo's BLR in the bright state spectra are $11,800 \pm 900$K and $11,900 \pm 2,400$K in 2022 and 2024 respectively, despite the highly variable accretion disc during the bright phase, shown by ZTF's optical light curves in Figure \ref{fig:photometry}. These values are of the order $10^4$K, expected for photoionisation equilibrium \citep{swiss_society_for_astrophysics_and_astronomy_agn_1990}, and fall within the range of BLR electron temperatures derived from Boltzmann plot analyses, which span roughly $7,000$–$37,000$K depending on the sample and method \citep{popovic_balmer_2003, popovic_broad_2006, mura_detailed_2007, ilic_analysis_2012}.

We observe a consistent BLR temperature between epochs. This may be explained by the fact that our optical continuum remains relatively constant between the two epochs: $L_{5100,\text{AGN}} = (2.3\pm0.1)\times10^{43}$ and $(2.4\pm0.1)\times10^{43} \text{erg s}^{-1}$ in 2022 and 2024, respectively. A strong correlation between the optical luminosity and BLR temperature was found using the same BP method for NGC 5548, a highly variable Seyfert 1 AGN, between 1998 and 2004 \citep{popovic_probing_2008}. The Balmer emission lines are primarily powered by photoionising UV photons from the accretion disc \citep{peterson_introduction_1997, osterbrock_astrophysics_2006}, however, we do not have access to UV observations of our target to probe this ionising continuum. In our spectra, we can only infer changes in the ionising continuum indirectly from variations in the blue tail of the optical accretion disc spectrum during the bright states. Further analysis of CLAGN with larger coverage of their spectral energy distributions is required to test the BLR temperature across a large population of changing-look sources.

\subsection{Implications of Using Single-epoch Black Hole Mass Estimates}
The extreme variability of broad emission in CLAGN can raise concerns about the applicability of single-epoch black hole mass estimates to all CLAGN states. In a virialised BLR, line width and luminosity are anti-correlated: when the continuum brightens, the BLR expands and lines narrow, and vice versa. This 'breathing' behaviour \citep{swiss_society_for_astrophysics_and_astronomy_agn_1990, korista_what_2004} means that different luminosity--width combinations can yield the same virial mass, as illustrated by the contours of constant mass in Figure \ref{fig:mass}. Testing whether such single-epoch estimates remain robust under the dramatic variability of CLAGN is therefore a key motivation of this work.

Between the dim state spectrum in 2004 and the bright states of 2022 and 2024, the decrease in FWHM of the H$\alpha$ broad line of $\sim1000 \text{km s}^{-1}$ and increase in luminosity by a factor of $3.9$, produce mass estimates that agree within the intrinsic scatter of the single-epoch relation from \citealt{bonta_estimating_2024} ($0.332$ dex). We therefore find no evidence from our analysis that CLAGN provide challenges to single-epoch mass relations.

This result implies that when variability is observed in the BLR through changes in broad emission lines, the virialised BLR assumption behind single-epoch approaches holds, at least within the uncertainties of the single epoch relation itself. This is particularly useful in the case of highly variable systems, such as changing-look AGN, where single-epoch measurements are typically taken from the brightest spectral states observed to maximise signal-to-noise of broad emission. Future monitoring of CLAGN through full transitions, especially from Type 1.0 to intermediate types (1.8/1.9), will be valuable to further test the robustness of single-epoch mass estimators. Not to mention, investigating additional broad lines commonly used for these estimations, such as \ion{Mg}{ii}, which show less variability in EVQs \citep{yang_discovery_2018}, may provide complementary mass constraints.

\section{Conclusion} \label{conclusion}

In this project, we investigate the changing-look AGN ZTF18abuamgo, following its turn-on event captured by ZTF in 2018, which triggered follow-up optical spectroscopic observations in 2022 and 2024. By comparing our new spectra to archival and pre-transient 6dF spectroscopy from 2004, we find that the broad Hydrogen Balmer lines have significantly brightened, and an optically blue tail, indicative of a luminous UV accretion disc, emerges during the bright state but disappears again two years later. Our main conclusions are summarised as follows:

\begin{enumerate}
    \item By measuring the line ratios between narrow [\ion{O}{iii}]$\lambda 5007$ and broad H$\beta$ emission, following the \citet{whittle_virial_1992} classification, we find the AGN to have transitioned from an intermediate Seyfert Type 1.5 in its 2004 dim-state to a Type 1.2 in bright-state 2022 and 2024 spectra. This spectral transition is less pronounced than that of many CLAGN studied before, often with broad emission lines that completely disappear. ZTF18abuamgo offers the opportunity to study intermediate CLAGN, which exhibit properties between typical AGN variability and the extreme tail in the AGN variability distribution.
    \item We estimate the changing-look transition timescale to be 4 years based on the maximum variability amplitude in the optical of $1$ magnitude between 2018 and 2022, similar to changing-look timescales studied from samples of CLAGN \citep{yang_discovery_2018, jana_investigating_2024}. Based on the spectroscopic epochs available, we obtain a conservative upper limit of 18 years for the timescale of the transition.
    \item We determine that the changing-look transition is driven by rapid changes in the accretion rate, classifying this source as a changing-state AGN. The presence of an optical blue tail in the 2022 bright-state spectrum from a UV-luminous accretion disc suggests a change in accretion rate, and the correlated increase in optical and MIR brightness argues against obscuration as a cause for the transient event. Additionally, the shape of the light curves disfavours a tidal disruption event scenario.
    \item We measure the accretion rate change by calculating the Eddington ratio from the optical continuum of the three spectral epochs. Using the optical continuum with strict upper and lower limits by utilising the dim-state continuum to constrain the brightness of the host contribution, we find the Eddington ratio increased from $0.032\pm0.005$ to $0.080\pm0.007$ in the bright state. These values are consistent with CLAGN identified by \citet{zeltyn_exploring_2024} in SDSS V, which show a bias towards Eddington ratios of a few per cent. The shift in Eddington ratio between spectral states in 2004 and 2022 or 2024 observations is also consistent with CLAGN studied by \citet{macleod_changing-look_2019, green_time_2022, lyu_wise_2022, jana_investigating_2024, dong_newly_2025}.
    \item For the first time, we have obtained temperature estimates for the BLR of a CLAGN using Boltzmann plots. We demonstrate that the temperature in ZTF18abuamgo's highly variable BLR is within the expected range of $1\times10^4$-$3\times10^4$ by using the Hydrogen Balmer line series present in the bright state spectra (2022 and 2024). The estimated electron temperatures we obtain are $11800\pm900$ K and $11900\pm2400$ K, respectively. The disappearance of the optically blue tail from the UV disc emission between these epochs suggests that the BLR maintains a constant temperature while the photoionising continuum varies.
    \item We test the efficacy of using single-epoch black hole mass measurements on a CLAGN that displays large variability on the broad H$\alpha$ and H$\beta$ lines. We find the mass of the SMBH to be $(5.04\pm0.38)\times10^7 M_{\odot}$ using H$\alpha$ measured in 2024 with the single-epoch black hole mass relations of \citet{bonta_estimating_2024}. We observe that the dim-state spectra, with fainter and larger line widths for H$\alpha$ and H$\beta$, consistently produce the lowest mass estimates with each single-epoch mass relation tested. However, due to the large intrinsic scatter associated with these relations of $\sim0.3$-$0.5$ dex, we demonstrate that for this CLAGN, the mass measurements are consistent across all epochs.
\end{enumerate}

\section*{Acknowledgements}
JC thanks the anonymous referee for their thoughtful feedback on this submission.
JC thanks Chris Frohmaier and Phil Wiseman for useful discussions and insights into transient astronomy, helping to overcome challenging parts of this work.
JC acknowledges the support of an STFC/UKRI PhD studentship reference 2890934 part of the project ST/Y509565/1.
This work was supported by the Science and Technology Facilities Council (STFC) of the UK Research and Innovation via grant reference ST/Y002644/1.

This publication also makes use of data products from NEOWISE, which is a project of the Jet Propulsion Laboratory/California Institute of Technology, funded by the Planetary Science Division of the National Aeronautics and Space Administration.
The data presented here were obtained [in part] with ALFOSC, which is provided by the Instituto de Astrofisica de Andalucia (IAA) under a joint agreement with the University of Copenhagen and NOT.
This work has made use of data from the Asteroid Terrestrial-impact Last Alert System (ATLAS) project. The Asteroid Terrestrial-impact Last Alert System (ATLAS) project is primarily funded to search for near earth asteroids through NASA grants NN12AR55G, 80NSSC18K0284, and 80NSSC18K1575; byproducts of the NEO search include images and catalogs from the survey area. 
This research has made use of the SVO Filter Profile Service "Carlos Rodrigo", funded by MCIN/AEI/10.13039/501100011033/ through grant PID2023-146210NB-I00.

Parts of this research were supported by the Australian Research Council Discovery Early Career Researcher Award (DECRA) through project number DE230101069.
Based in part on data acquired at the ANU 2.3-metre telescope, under proposal ID: 2425062. The automation of the telescope was made possible through an initial grant provided by the Centre of Gravitational Astrophysics and the Research School of Astronomy and Astrophysics at the Australian National University and through a grant provided by the Australian Research Council through LE230100063. The Lens proposal system is maintained by the AAO Research Data \& Software team as part of the Data Central Science Platform. We acknowledge the traditional custodians of the land on which the telescope stands, the Gamilaraay people, and pay our respects to elders past and present.

\textit{Software:} Astropy \citep{the_astropy_collaboration_astropy_2013, the_astropy_collaboration_astropy_2018, the_astropy_collaboration_astropy_2022}, PyQSOFit \citep{guo_pyqsofit_2018, shen_sloan_2019, ren_prior-informed_2024}

\section*{Data Availability}

Data generated in this research will be shared on reasonable request to the corresponding author.


\bibliographystyle{mnras}
\input{mnras_template.bbl}


\bsp	
\label{lastpage}
\end{document}